\documentclass[%
 aip,
 amsmath,amssymb,
 reprint,%
]{revtex4-1}

\usepackage{graphicx}
\usepackage{dcolumn}
\usepackage{bm}

\usepackage[cal=boondox]{mathalfa}

\usepackage[utf8]{inputenc}
\usepackage[T1]{fontenc}
\usepackage{mathptmx}

\DeclareMathAlphabet\mathbfcal{OMS}{cmsy}{b}{n}

\DeclareMathOperator{\sn}{sn}
\DeclareMathOperator{\cn}{cn}

\begin{document}

\preprint{AIP/123-QED}

\title[Dispersion Properties, Nonlinear Waves and Birefringence ...]{Dispersion Properties, Nonlinear Waves and Birefringence in Classical Nonlinear Electrodynamics}

\author{Stephan I. Tzenov}
\email{stephan.tzenov@eli-np.ro}
\affiliation{Extreme Light Infrastructure - Nuclear Physics, 191014 Magurele, jud. Ilfov County, 30 Reactorului, Romania}

\author{Klaus M. Spohr}%
\affiliation{Extreme Light Infrastructure - Nuclear Physics, 191014 Magurele, jud. Ilfov County, 30 Reactorului, Romania}%

\author{Kazuo A. Tanaka}%
\affiliation{Extreme Light Infrastructure - Nuclear Physics, 191014 Magurele, jud. Ilfov County, 30 Reactorului, Romania}%

\date{\today}

\begin{abstract}
Using the very basic physics principles, we have studied the implications of quantum corrections to classical electrodynamics and the propagation of electromagnetic waves and pulses. 

The initial nonlinear wave equation for the electromagnetic vector potential is solved perturbatively about the known exact plane wave solution in both the free vacuum case, as well as when a constant magnetic field is applied. A nonlinear wave equation with nonzero convective part for the (relatively) slowly varying amplitude of the first-order perturbation has been derived. This equation governs the propagation of electromagnetic waves with a reduced speed of light, where the reduction is roughly proportional to the intensity of the initial pumping plane wave. A system of coupled nonlinear wave equations for the two slowly varying amplitudes of the first-order perturbation, which describe the two polarization states, has been obtained for the case of constant magnetic field background. 

Further, the slowly varying wave amplitude behavior is shown to be similar to that of a cnoidal wave, known to describe surface gravity waves in shallow water. It has been demonstrated that the two wave modes describing the two polarization states are independent, and they propagate at different wave frequencies. This effect is usually called nonlinear birefringence.
\end{abstract}

\pacs{12.20.Ds, 52.35.Mw, 95.30.Cq}
\keywords{ nonlinear waves, wave propagation, nonlinear birefringence}
\maketitle

%

\section{\label{sec:intro}Introduction} 

One of the most fundamental and important features of classical electrodynamics is the fact that macroscopically it is a linear theory. This property of Maxwell electrodynamics described in every standard textbook finds remarkable validation in physics experiments and relevant applications almost daily. Put otherwise in a formal language, the linearity of classical electrodynamics sounds like this: any superposition of two or more electromagnetic fields obeying Maxwell's equations satisfies the latter as well\cite{Zangwill}. 

At the subatomic level, however, deviations from the principle of linear superposition can be expected. The standard simplification commonly used in theoretical physics consists of the representation of a charged particle as a localized distribution of charge. Unfortunately, this leads to the infinite growth of its electromagnetic energy with the decrease of the localization dimensions, thus approaching a point-like distribution. To avoid infinite self-energies of point particles, it is natural to speculate that a particular field strength saturation (upper bound) exists. One well-known example of such nonlinear theories developed in the past is the theory of Born and Infeld\cite{Born}. Classical electrodynamics cannot describe the interaction between two electromagnetic waves, whereas in quantum electrodynamics, such a scattering of light by light is possible. The two incident plane waves with wave vectors ${\bf k}_1$ and ${\bf k}_2$ do not merely add coherently, as predicted by linear superposition, but interact and (with a small probability) transform into two different plane waves with corresponding wave vectors ${\bf k}_3$ and ${\bf k}_4$. These results were first obtained by Euler and Kockel\cite{EulKock} in 1935 and further elaborated by Heisenberg and Euler\cite{EulHeisen} in 1936.
Interestingly enough in both Born-Infeld and Heisenberg-Euler theories of nonlinear electrodynamics in the weak-field limit, the electric polarization ${\bf P}$ and the magnetization ${\bf M}$ vectors are of third-order (up to leading terms) in the electric ${\bf E}$ and the magnetic ${\bf B}$ field. This implies that the polarized quantum vacuum can be regarded as a nonlinear medium of Kerr-type. In the present article, we shall investigate the properties of the nonlinear Heisenberg-Euler electrodynamics in the weak-field limit. 

It seems that the linear dispersion properties of photons in a constant, or weakly varying electromagnetic background fields have been first studied in the early 1960s\cite{Birula}. In order to take into account external field variations for space and time, the weak-field Heisenberg-Euler Lagrangian must be modified by including a term containing derivatives of the electromagnetic field tensor\cite{Mamaev,Rozanov}. The systematic consideration and discussion of dispersion effects in the case of strong electromagnetic fields have been performed in Ref. \citenum{Dittrich}.

The analysis of a number of applications, in which nonlinear collective effects among photons may play a dominant role has been initiated by N.N. Rozanov\cite{Rozanov,Rozanov1}. He considered the perpendicular propagation of high-intensity laser pulses traveling in externally applied electric and magnetic cross fields. By choosing an initial pumping laser pulse polarized in the direction of the background fields, Rozanov obtained a nonlinear Schr{\"o}dinger equation for the slowly varying envelope of the perturbation\cite{Rozanov}. Soljacic and Segev\cite{Segev} studied the case of a "beam" resulting from the superposition of two carrier plane waves, modified by a slowly varying envelope in the horizontal direction. They derived a nonlinear Schr{\"o}dinger equation governing the dynamics of the "beam" envelope due to the crossing laser waves. Brodin et al.\cite{Brodin} considered the propagation of a single TE-mode with weakly modulated amplitude between two parallel conducting planes, and as a result, came up with a similar nonlinear Schr{\"o}dinger equation for the slowly varying mode envelope. 

Compared with the usual properties of various media widely used in nonlinear optics, the vacuum is characterized by both the nonlinear electric polarization and the nonlinear magnetization simultaneously. In the weak-field limit, these nonlinearities give rise to three and four-wave mixing, which have been studied in Refs. \citenum{Rozanov1}, \citenum{Brodin1} and \citenum{Brodin2}. 

In the present article, we adopt a different approach, as compared to other works dedicated to nonlinear wave phenomena in polarized vacuum existing in the literature. It is a simple and rather intuitive one, and is widely used in finite and infinite degree-of-freedom dynamical systems. The perturbation method we utilize here is quite straightforward, and as a starting point, it uses the availability of an exact solution to the initial nonlinear wave equations describing the properties of the quantum vacuum. The perturbation expansion about the exact solution of the underlying equations is then built up to third-order, followed by the renormalization of secular terms. Thus, the relevant dynamics is being split into two parts: the first one describing the fast wave oscillations, and a second one exhibiting the nonlinear behavior of specific wave amplitudes on much slower spatial and time scales. 

The article is organized as follows. In the next Section, we present some basics of the weak-field nonlinear electrodynamics of Heisenberg-Euler type. In Section \ref{sec:nlindisp}, we study the free (from external electromagnetic field) quantum vacuum. Small perturbation about the exact plane wave solution of the nonlinear wave equation for the electromagnetic vector potential has been analysed. Employing the Renormalization Group (RG) method\cite{Chen,TzenovBOOK}, we derive a nonlinear wave equation for the (relatively) slowly varying amplitude of the first-order perturbation. An intriguing property of this equation is that it governs the propagation of electromagnetic waves with a reduced speed of light. In Section \ref{sec:nlindispmf}, we consider the case where an external magnetic field is applied. The latter introduces spatial anisotropy with all the ensuing consequences. Similar to the free vacuum case, the underlying wave equation for the electromagnetic vector potential possesses an exact solution in the form of a plane wave with constant amplitude. For the analysis carried out in the present article, we choose the circularly polarized exact solution and repeat the RG procedure in a similar manner as has been done in Section \ref{sec:nlindisp}. As a result, we obtain a system of coupled nonlinear wave equations for the two slowly varying amplitudes of the first-order perturbation, which describe the two polarization states. In Section \ref{sec:soliton}, we present the stationary (traveling) wave solutions of the nonlinear wave equation, thus obtained for the case of a free polarized vacuum. The slowly varying wave amplitude behavior is shown to be similar to that of a cnoidal wave, characterized by a near-periodic swell in shallow water. Finally, in Section \ref{sec:conclude} we draw some conclusions. 

\section{\label{sec:basic}Theoretical Model and Basic Equations}

In Quantum Electrodynamics, photon-photon scattering is known to be a second-order effect in terms of the fine-structure constant $\alpha$. For constant or weakly varying fields it can be formulated in standard notation by using the Lagrangian (Euler-Heisenberg Lagrangian density\cite{Schwinger}) or the Hamiltonian approach\cite{Vollick} in classical field theory. The effects of the quantum electrodynamics vacuum polarization and magnetization introduce (third-order) nonlinearity, such that vacuum itself can be considered a nonlinear medium with appropriate constitutive relations 
\begin{equation}
{\bf D} = \epsilon_0 {\bf E} + {\bf P}, \qquad \qquad {\bf H} = {\frac {1} {\mu_0}} {\bf B} - {\bf M}. \label{ElDisplInduc}
\end{equation}
Here $\epsilon_0$ and $\mu_0$ are the electric permittivity and the magnetic permeability of vacuum ${\left( \epsilon_0 \mu_0 = 1 / c^2 \right)}$, respectively, ${\bf D}$ and ${\bf H}$ are the electric displacement vector and the intensity of the magnetic field, respectively, while ${\bf P}$ is the electric polarization and ${\bf M}$ is the magnetization. The latter two quantities are third-order in ${\bf E}$ and ${\bf B}$
\begin{equation}
{\bf P} = 2 \kappa \epsilon_0^2 {\left[ 2 {\left( E^2 - c^2 B^2 \right)} {\bf E} + 7 c^2 {\left( {\bf E} \cdot {\bf B} \right)} {\bf B} \right]}, \label{Polarization}
\end{equation}
\begin{equation}
{\bf M} = 2 \kappa \epsilon_0^2 c^2 {\left[ - 2 {\left( E^2 - c^2 B^2 \right)} {\bf B} + 7 {\left( {\bf E} \cdot {\bf B} \right)} {\bf E} \right]}, \label{Magnetization}
\end{equation}
where 
\begin{equation}
\kappa = {\frac {2 \alpha^2 \hbar^3} {45 m_e^4 c^5}} = {\frac {\alpha} {90 \pi}} {\frac {1} {\epsilon_0 E_S^2}} \approx {\frac {1} {3 \times 10^{29} \; {\rm J/m^3}}}. \label{Kappa}
\end{equation}
Here $m_e$ is the electron rest mass, and $E_S = m_e^2 c^3 / (e \hbar) \approx 1.32 \times 10^{18} \; {\rm V / m}$, is the Schwinger limit above which the electromagnetic field is expected to become nonlinear. Equations (\ref{Polarization}) and (\ref{Magnetization}) indicate that the nonlinear corrections take the same form as in nonlinear optics, where the material properties of optical fibers for example, give rise to cubic nonlinear terms in Maxwell’s equations, so-called Kerr effect. The essential difference however, is that quantum vacuum polarization leads to nonlinearities in both the electric polarization and the magnetization. Note that the effective self-interaction term is proportional to the fine structure constant squared, which implies that field strengths must reach values close to the Schwinger limit until these effects become important. 

The macroscopic Maxwell equations taking into account the vacuum polarization and magnetization can be written as 
\begin{equation}
{\boldsymbol{\nabla}} \cdot {\bf D} = \varrho_f, \qquad \qquad {\boldsymbol{\nabla}} \cdot {\bf B} = 0, \label{Maxwell1}
\end{equation}
\begin{equation}
{\boldsymbol{\nabla}} \times {\bf E} = - \partial_t {\bf B}, \qquad \qquad {\boldsymbol{\nabla}} \times {\bf H} - \partial_t {\bf D} = {\bf J}_f, \label{Maxwell2}
\end{equation}
where $\varrho_f$ and ${\bf J}_f$ is the free charge and current densities, respectively. Manipulating Eqs. (\ref{Maxwell1}) and (\ref{Maxwell2}) in an obvious manner with due account of Eq. (\ref{ElDisplInduc}), we obtain 
\begin{eqnarray}
{\Box} {\bf E} = \mu_0 && {\left[ \partial_t^2 {\bf P} - c^2 {\boldsymbol{\nabla}} {\left( {\boldsymbol{\nabla}} \cdot {\bf P} \right)} + \partial_t {\left( {\boldsymbol{\nabla}} \times {\bf M} \right)} \right]} \nonumber 
\\ 
&& + \mu_0 {\left( \partial_t {\bf J}_f + c^2 {\boldsymbol{\nabla}} \varrho_f \right)}, \label{WaveEqE}
\end{eqnarray}
\begin{eqnarray}
{\Box} {\bf B} = - \mu_0 && {\left[ - {\boldsymbol{\nabla}}^2 {\bf M} + {\boldsymbol{\nabla}} {\left( {\boldsymbol{\nabla}} \cdot {\bf M} \right)} + \partial_t {\left( {\boldsymbol{\nabla}} \times {\bf P} \right)} \right]} \nonumber 
\\ 
&& - \mu_0 {\boldsymbol{\nabla}} \times {\bf J}_f, \label{WaveEqB}
\end{eqnarray}
where ${\Box} = {\boldsymbol{\nabla}}^2 - {\left( 1 / c^2 \right)} \partial_t^2$ is the well-known d'Alembert operator. These are the nonlinear generalizations of the classical wave equations for the electric and the magnetic field, often serving as a starting point in the description of a variety of nonlinear phenomena in quantum vacuum.

In what follows, however, more convenient will be the alternative formulation of nonlinear electrodynamics in the weak-field limit. Since the second of Eqs. (\ref{Maxwell1}) and the first of Eqs. (\ref{Maxwell2}) remain unaltered as compared to the microscopic electrodynamics, they allow the introduction of the scalar $\Phi$ and the vector ${\bf A}$ potentials defined in the standard manner 
\begin{equation}
{\bf E} = - {\boldsymbol{\nabla}} \Phi - \partial_t {\bf A}, \qquad \qquad {\bf B} = {\boldsymbol{\nabla}} \times {\bf A}. \label{ScalVecPot}
\end{equation}
Manipulating the remaining two of the Maxwell equations and taking into account the definition (\ref{ElDisplInduc}) for the electric displacement vector and the intensity of the magnetic field, we obtain 
\begin{widetext}
\begin{equation}
{\Box} {\bf A} - {\boldsymbol{\nabla}} {\left( {\frac {1} {c^2}} \partial_t \Phi + {\boldsymbol{\nabla}} \cdot {\bf A} \right)} = - \mu_0 {\left( \partial_t {\bf P} + {\boldsymbol{\nabla}} \times {\bf M} \right)} - \mu_0 {\bf J}_f, \label{WaveEqVecPot}
\end{equation}
\end{widetext}
\begin{equation}
{\boldsymbol{\nabla}}^2 \Phi + \partial_t {\left( {\boldsymbol{\nabla}} \cdot {\bf A} \right)} = {\frac {1} {\epsilon_0}} {\boldsymbol{\nabla}} \cdot {\bf P} - {\frac {\varrho_f} {\epsilon_0}}. \label{WaveEqScalPot}
\end{equation}
In addition to the nonlinear wave equations above, one can use the Lorentz ${\left[ {\left( 1 / c^2 \right)} \partial_t \Phi + {\boldsymbol{\nabla}} \cdot {\bf A} = 0 \right]}$ or the Coulomb ${\left( {\boldsymbol{\nabla}} \cdot {\bf A} = 0 \right)}$ gauge condition. 

Equations (\ref{WaveEqVecPot}) and (\ref{WaveEqScalPot}) together with Eqs. (\ref{Polarization}) and (\ref{Magnetization}) describe the nonlinear wave properties of polarized vacuum in the weak-field limit. The case, where free electric charges $\varrho_f$ and currents ${\bf J}_f$ are absent will be the starting point for our subsequent analysis. 

\section{\label{sec:nlindisp}Nonlinear Waves in Free Polarized Vacuum and the Nonlinear Amplitude Equation} 

We assume a field configuration as follows 
\begin{equation}
{\bf E} = {\left( 0, \; - \partial_t A, \; 0 \right)}, \qquad {\bf B} = {\left( - \partial_z A, \; 0, \; \partial_x A \right)}, \label{FieldConfig}
\end{equation}
which is determined by a vector potential 
\begin{equation}
{\bf A} {\left( x, z; t \right)} = {\left[ 0, \; A {\left( x, z; t \right)}, \; 0 \right]}. \label{VecPotConfig}
\end{equation}
Since ${\bf E} \cdot {\bf B} = 0$ in this configuration, the second terms in Eqs. (\ref{Polarization}) and (\ref{Magnetization}) depending on the scalar product of the electric field ${\bf E}$ and the magnetic field ${\bf B}$ vanish. As far as components are concerned, obviously, the vacuum polarization ${\bf P}$ and the vacuum magnetization ${\bf M}$ follow the pattern of ${\bf E}$ and ${\bf B}$
\begin{equation}
P_y = - 4 \kappa \epsilon_0^2 c^2 {\cal F} \partial_t A, \qquad {\bf M} = - 4 \kappa \epsilon_0^2 c^4 {\cal F} {\boldsymbol{\nabla}} \times {\bf A}, \label{PolarizPM}
\end{equation}
respectively. Here 
\begin{equation}
{\cal F} = {\frac {1} {c^2}} {\left( \partial_t A \right)}^2 - {\left( {\boldsymbol{\nabla}} A \right)}^2. \label{FFunction}
\end{equation}
It is convenient to use the Coulomb gauge, in which the scalar potential $\Phi$ vanishes identically. 

It can be verified in a straightforward manner that the nonlinear wave equation (\ref{WaveEqVecPot}) for the vector potential can be rewritten as 
\begin{equation}
{\Box} A = 4 \kappa \epsilon_0 \partial_t {\left( {\cal F} \partial_t A \right)} - 4 \kappa \epsilon_0 c^2 {\left( {\boldsymbol{\nabla}} A \cdot {\boldsymbol{\nabla}} {\cal F} + {\cal F} {\boldsymbol{\nabla}}^2 A \right)}. \label{BasEquatVecPot}
\end{equation}
Note that a plane wave of the form 
\begin{equation}
A_0 = {\cal C} {\rm e}^{i \varphi_0} + {\cal C}^{\ast} {\rm e}^{- i \varphi_0}, \label{PlaneWaveSol}
\end{equation}
where ${\cal C}$ is a constant complex amplitude, $\varphi_0 = {\bf k}_0 \cdot {\bf x} - \omega_0 t$ is the wave phase, and the wave frequency $\omega_0$ and the wave number ${\bf k}_0$ satisfy the dispersion relation 
\begin{equation}
\omega_0^2 = c^2 {\left( k_{0x}^2 +k_{0z}^2 \right)} = c^2 k_0^2, \label{LinearDispers}
\end{equation}
is an exact solution of Eq. (\ref{BasEquatVecPot}). This property is not surprising -- it immediately follows from the important relation ${\cal F}_0 = {\cal F} {\left( A_0 \right)} = 0$, which can be checked by direct substitution. Put another way, the plane wave minimizes the effective Heisenberg-Euler action.

Following the standard procedure of the renormalization group method, we represent $A {\left( x, z; t \right)}$ as a perturbation expansion
\begin{equation}
A {\left( x, z; t \right)} = \sum \limits_{n=0}^{\infty} \epsilon^n A_n {\left( x, z; t \right)}, \label{RGExpans}
\end{equation}
in the formal small parameter $\epsilon$. The next step consists in expanding Eq. (\ref{BasEquatVecPot}) in the small parameter $\epsilon$, and obtaining its naive perturbation solution order by order. 

\subsection{\label{subsec:forder}First-Order} 

The first-order vector potential $A_1$ obeys the equation 
\begin{equation}
{\widehat{\bf L}}_1 A_1 = - \alpha {\left[ {\rm e}^{2i {\left( \varphi_0 + \delta \right)}} + {\rm e}^{-2i {\left( \varphi_0 + \delta \right)}} \right]} {\widehat{\bf D}}_1^2 A_1, \label{LinearizedFO}
\end{equation}
where 
\begin{equation}
{\widehat{\bf D}}_1 = {\frac {\omega_0} {c}} \partial_t + c {\bf k}_0 \cdot {\boldsymbol{\nabla}}, \qquad \quad {\widehat{\bf L}}_1 = {\Box} - 2 \alpha {\widehat{\bf D}}_1^2, \label{BasOperat}
\end{equation}
\begin{equation}
\alpha = 8 \kappa \epsilon_0 {\left| {\cal C} \right|}^2, \label{AlphaCoeff}
\end{equation}
and $\delta$ is the phase of the constant complex amplitude ${\cal C}$. The right-hand-side of Eq. (\ref{LinearizedFO}) is fast oscillating, and can be averaged away. Thus, we obtain the first-order solution 
\begin{equation}
A_1 = {\cal A} {\rm e}^{i \varphi_1} + {\cal A}^{\ast} {\rm e}^{- i \varphi_1}. \label{FirstOrderSol}
\end{equation}
Here, ${\cal A}$ is an additional constant (for the present moment in time) complex amplitude and $\varphi_1 = {\bf k} \cdot {\bf x} - \omega t$ is the wave phase. In addition, the wave frequency 
\begin{widetext}
\begin{equation}
\omega {\left( {\bf k} \right)} = {\frac {1} {1 + 2 \alpha \omega_0^2}} {\left[ 2 \alpha c^2 \omega_0 {\bf k}_0 \cdot {\bf k} + {\sqrt{c^2 k^2 {\left( 1 + 2 \alpha \omega_0^2 \right)} - 2 \alpha c^4 {\left( {\bf k}_0 \cdot {\bf k} \right)}^2}} \right]}. \label{OmegaFreq}
\end{equation}
\end{widetext}
is a solution to the dispersion relation 
\begin{equation}
{\cal D} {\left( {\bf k}, \omega \right)} = {\frac {\omega^2} {c^2}} - k^2 + 2 \alpha {\left( {\frac {\omega_0 \omega} {c}} - c {\bf k}_0 \cdot {\bf k} \right)}^2 = 0. \label{NonlinDisper}
\end{equation}

An important comment is now in order. Apart from the solution (\ref{FirstOrderSol}), the first-order equation (\ref{LinearizedFO}) possesses a solution of the form (\ref{PlaneWaveSol}), which has been disregarded, since it is already contained in the exact plane wave solution of the initial nonlinear wave equation. Furthermore, Eq. (\ref{LinearizedFO}) can be regarded as a wave analog of the Mathieu equation. Perturbation analysis shows that the fast oscillating term proportional to $\cos 2 {\left( \varphi + \delta \right)}$ does not give rise to additional higher-order contribution in terms of the polarization parameter $\kappa$. In other words, the neglection (or equivalently, the averaging procedure) of the fast oscillating term is justified up to (at least) second-order in perturbation theory as applied to the Mathieu equation (\ref{LinearizedFO}). Further technical details can be found in Appendix \ref{sec:mathieu}. 

\subsection{\label{subsec:sorder}second-order} 

The second-order perturbation equation for vector potential $A_2$ reads as 
\begin{eqnarray}
{\Box} A_2 = && 4 \kappa \epsilon_0 \partial_t {\left( {\cal F}_2 \partial_t A_0 + {\cal F}_1 \partial_t A_1 \right)} - 4 \kappa \epsilon_0 c^2 {\left( {\boldsymbol{\nabla}} A_0 \cdot {\boldsymbol{\nabla}} {\cal F}_2 \right.} \nonumber
\\ 
&& {\left. + {\boldsymbol{\nabla}} A_1 \cdot {\boldsymbol{\nabla}} {\cal F}_1 + {\cal F}_2 {\boldsymbol{\nabla}}^2 A_0 + {\cal F}_1 {\boldsymbol{\nabla}}^2 A_1 \right)}
. \label{SecOrdPertA2}
\end{eqnarray}
Substituting Eqs. (\ref{PlaneWaveSol}) and (\ref{FirstOrderSol}) into the right-hand-side of Eq. (\ref{SecOrdPertA2}), we rewrite the latter in a detailed form as follows 
\begin{widetext}
\begin{equation}
{\widehat{\bf L}}_1 A_2 + 2 \alpha \cos 2 {\left( \varphi_0 + \delta \right)} {\widehat{\bf D}}_1^2 A_2 = 8 \kappa \epsilon_0 c^2 {\Box}_{0k} {\left[ {\left( {\Box}_{0k} + 3 {\Box}_k \right)} {\cal C} {\cal A}^2 {\rm e}^{i {\left( 2 \varphi_1 + \varphi_0 \right)}} + {\left( {\Box}_{0k} - 3 {\Box}_k \right)} {\cal C}^{\ast} {\cal A}^2 {\rm e}^{i {\left( 2 \varphi_1 - \varphi_0 \right)}} + c.c. \right]}, \label{SecOrdPertEqu}
\end{equation}
\end{widetext}
where 
\begin{equation}
{\Box}_k = {\frac {\omega^2} {c^2}} - k^2, \qquad \qquad {\Box}_{0k} = {\frac {\omega_0 \omega} {c^2}} - {\bf k}_0 \cdot {\bf k}. \label{BoxNotat}
\end{equation}
The solution of the second-order perturbation equation (\ref{SecOrdPertEqu}) can be written as 
\begin{widetext}
\begin{equation}
A_2 = 2 \kappa \epsilon_0 c^2 {\left[ {\left( {\Box}_{0k} + 3 {\Box}_k \right)} {\cal C} {\cal A}^2 {\rm e}^{i {\left( 2 \varphi_1 + \varphi_0 \right)}} - {\left( {\Box}_{0k} - 3 {\Box}_k \right)} {\cal C}^{\ast} {\cal A}^2 {\rm e}^{i {\left( 2 \varphi_1 - \varphi_0 \right)}}  + c.c. \right]}. \label{SecondOrderSol}
\end{equation}
\end{widetext}
Note that the second-order vector potential $A_2$ includes terms with combination phases of the form ${\rm e}^{i {\left( 2 \varphi_1 \pm \varphi_0 \right)}}$. They describe the three wave mixing and are giving rise to secular terms as will be shown below. 

\subsection{\label{subsec:torder}third-order. Derivation of the Nonlinear Amplitude Equation} 

The third-order perturbation equation for vector potential $A_3$ reads as 
\begin{widetext}
\begin{eqnarray}
{\Box} A_3 = 4 \kappa \epsilon_0 \partial_t {\left( {\cal F}_3 \partial_t A_0 + {\cal F}_2 \partial_t A_1 + {\cal F}_1 \partial_t A_2 \right)} \nonumber 
\\ 
- 4 \kappa \epsilon_0 c^2 {\left( {\boldsymbol{\nabla}} A_0 \cdot {\boldsymbol{\nabla}} {\cal F}_3 + {\boldsymbol{\nabla}} A_1 \cdot {\boldsymbol{\nabla}} {\cal F}_2 + {\boldsymbol{\nabla}} A_2 \cdot {\boldsymbol{\nabla}} {\cal F}_1 + {\cal F}_3 {\boldsymbol{\nabla}}^2 A_0 + {\cal F}_2 {\boldsymbol{\nabla}}^2 A_1 + {\cal F}_1 {\boldsymbol{\nabla}}^2 A_2 \right)}
. \label{ThirdOrdPertA3}
\end{eqnarray}
\end{widetext}
Unlike the second-order, where the perturbation equation possesses a unique regular solution, there are two types of terms on the right-hand-side of the third-order equation (\ref{ThirdOrdPertA3}). The first type comprises a collection of secular (resonant) terms, which follow the pattern of the basic wave mode in linear approximation (proportional to ${\rm e}^{\pm i \varphi_1}$). Such terms, which appear in higher orders as well, would provide a divergent counterpart in the naive perturbation solution. Thus, they must be renormalized by an elegant procedure described below. The rest of the terms contribute to the regular solution of the third-order perturbation equations, involving higher harmonics and/or higher order harmonic combinations of the modes proportional to ${\rm e}^{i \varphi_0}$ and ${\rm e}^{i \varphi_1}$. Omitting straightforwardly reproducible calculation's details, we write down the resonant part of Eq. (\ref{ThirdOrdPertA3}) 
\begin{eqnarray}
{\widehat{\bf L}}_1 A_3 + 2 \alpha \cos 2 {\left( \varphi_0 + \delta \right)} {\widehat{\bf D}}_1^2 A_3 \nonumber 
\\ 
= - 36 \kappa \epsilon_0 c^2 {\Box}_k^2 {\left| {\cal A} \right|}^2 {\cal A} {\rm e}^{i \varphi_1} + c.c., \label{ThiOrdPertEqu}
\end{eqnarray}
The above equation possesses an exact solution 
\begin{equation}
A_3 {\left( x, z; t \right)} = {\cal P}_3 {\left( x, z; t \right)} {\rm e}^{i \varphi_1}, \label{ThirdOrderSol}
\end{equation}
where the amplitude ${\cal P}_3 {\left( x, z; t \right)}$ satisfies the equation 
\begin{eqnarray}
{\left[ {\widehat{\bf L}}_1 + i {\left( \partial_{\omega} {\cal D} \right)} \partial_t - i {\left( {\boldsymbol{\nabla}}_k {\cal D} \right)} \cdot {\boldsymbol{\nabla}} \right]} {\cal P}_3 \nonumber
\\ 
= - 36 \kappa \epsilon_0 c^2 {\Box}_k^2 {\left| {\cal A} \right|}^2 {\cal A}. \label{AmpEquP3}
\end{eqnarray}
Here ${\cal D} {\left( {\bf k}, \omega \right)}$ is the dispersion function defined by Eq. (\ref{NonlinDisper}).

We follow an elegant approach, known as the proto RG operator scheme\cite{OonoNoz,NOS,Shiwa}, which has been proposed in the early 2000s to free as much as possible the standard RG theoretical reduction from the necessity of explicit (in the majority of cases, rather cumbersome) calculation of secular terms [see e.g. Refs. \citenum{Chen} and \citenum{TzenovBOOK}]. Thus, we finally arrive at the sought for nonlinear amplitude equation 
\begin{eqnarray}
{\left[ {\widehat{\bf L}}_1 + i {\left( \partial_{\omega} {\cal D} \right)} \partial_t - i {\left( {\boldsymbol{\nabla}}_k {\cal D} \right)} \cdot {\boldsymbol{\nabla}} \right]} {\cal A} \nonumber
\\ 
= - 36 \kappa \epsilon_0 c^2 {\Box}_k^2 {\left| {\cal A} \right|}^2 {\cal A}. \label{NonlinAmpEqu}
\end{eqnarray}
It governs the relatively slow dynamics of the wave envelope ${\cal A}$ and describes the formation of nonlinear waves and coherent structures. As it will be shown in Section \ref{sec:soliton} the equation just derived is a nonlinear wave equation with a characteristic reduced light velocity. 

\section{\label{sec:nlindispmf}Nonlinear Dispersion Relation in Polarized Vacuum with Externally Applied Magnetic Field} 

In this Section, we shall analyze the case, where a constant external magnetic field ${\bf B}_0 = {\left( B_0, 0, 0 \right)}$ is applied along the $x$-axis. Since the applied magnetic field introduces a special type of spatial anisotropy, we assume that the induced electromagnetic waves propagate in the axial $x$-direction only. The electromagnetic field configuration can be specified as follows 
\begin{equation}
{\bf E} = {\left( 0, \; - \partial_t A_y, \; - \partial_t A_z \right)}, \qquad {\bf B} = {\left( B_0, \; - \partial_x A_z, \; \partial_x A_y \right)}, \label{FieldConfigMF}
\end{equation}
which is defined by an electromagnetic vector potential of the form 
\begin{equation}
{\bf A} {\left( x; t \right)} = {\left[ 0, \; A_y {\left( x; t \right)}, \; A_z {\left( x; t \right)} \right]}. \label{VecPotConfigMF}
\end{equation}
Since for plane waves this configuration yields ${\bf E} \cdot {\bf B} = 0$, the second terms in Eqs. (\ref{Polarization}) and (\ref{Magnetization}) depending on the scalar product of the electric field ${\bf E}$ and the magnetic field ${\bf B}$ vanish correspondingly. 

It will prove convenient for the subsequent exposition to introduce a new field variable according to the relation 
\begin{equation}
{\cal A} {\left( x; t \right)} = A_y {\left( x; t \right)} + i A_z {\left( x; t \right)}. \label{ComplexVecPot}
\end{equation}
Obviously, 
\begin{equation}
E^2 = \partial_t {\cal A} {\left( \partial_t {\cal A}^{\ast} \right)}, \qquad B^2 = B_0^2 + \partial_x {\cal A} {\left( \partial_x {\cal A}^{\ast} \right)}. \label{ElMagSquare}
\end{equation}
As far as components are concerned, similarly to the preceding Section the vacuum polarization ${\bf P}$ and the vacuum magnetization ${\bf M}$ follow the vectorial pattern of ${\bf E}$ and ${\bf B}$
\begin{widetext}
\begin{equation}
P_x = 0, \quad \qquad P_y = - 4 \kappa \epsilon_0^2 c^2 {\cal G} \partial_t A_y, \quad \qquad P_z = - 4 \kappa \epsilon_0^2 c^2 {\cal G} \partial_t A_z, \label{PolarizMF}
\end{equation}
\begin{equation}
M_x = - 4 \kappa \epsilon_0^2 c^4 {\cal G} B_0, \quad \qquad M_y = 4 \kappa \epsilon_0^2 c^4 {\cal G} \partial_x A_z, \quad \qquad M_z = - 4 \kappa \epsilon_0^2 c^4 {\cal G} \partial_x A_y, \label{MagnetizMF}
\end{equation}
\end{widetext}
respectively. Here 
\begin{equation}
{\cal G} = {\frac {1} {c^2}} \partial_t {\cal A} {\left( \partial_t {\cal A}^{\ast} \right)} - \partial_x {\cal A} {\left( \partial_x {\cal A}^{\ast} \right)} - B_0^2. \label{GFunction}
\end{equation}
As it was done in in the case of a free polarized quantum vacuum, we again use the Coulomb gauge, in which the scalar potential $\Phi$ vanishes identically. 

It can be verified in a straightforward manner that the nonlinear wave equation (\ref{WaveEqVecPot}) for the vector complex potential can be rewritten as 
\begin{equation}
{\Box} {\cal A} = 4 \kappa \epsilon_0 \partial_t {\left( {\cal G} \partial_t {\cal A} \right)} - 4 \kappa \epsilon_0 c^2 \partial_x {\left( {\cal G} \partial_x {\cal A} \right)}. \label{BasEquatVecPotMF}
\end{equation}
Again, it can be checked by direct substitution that a plane wave of the form 
\begin{equation}
{\cal A}_0 = {\cal E} {\left( {\rm e}^{i \varphi_0} + {\cal P} {\rm e}^{-i \varphi_0} \right)}, \label{PlaneWaveSolMF}
\end{equation}
where ${\cal E}$ is a constant complex amplitude, $\varphi_0 = k_0 x - \omega_0 t$ is the wave phase, and the wave frequency $\omega_0$ and the wave number $k_0$ satisfy the dispersion relation $\omega_0^2 = c^2 k_0^2$, is an exact solution of Eq. (\ref{BasEquatVecPotMF}). In addition, ${\cal P}$ is the polarization parameter, taking into account various plane wave polarizations.  The assertion that the plane wave (\ref{PlaneWaveSolMF}) is an exact solution immediately follows from the important property ${\cal G}_0 = {\cal G} {\left( {\cal A}_0 \right)} = - B_0^2$. For the sake of simplicity we consider here the case of circular polarization, that is ${\cal P} = 0$. 

Similar to the preceding Section, we represent ${\cal A}$ as a perturbation expansion
\begin{equation}
{\cal A} {\left( x; t \right)} = \sum \limits_{n=0}^{\infty} \epsilon^n {\cal A}_n {\left( x; t \right)}, \label{RGExpansMF}
\end{equation}
in the formal small parameter $\epsilon$. The first-order vector potential ${\cal A}_1$ obeys now the equation 
\begin{equation}
{\widehat{\bf L}}_M {\cal A}_1 = - \beta {\rm e}^{2i {\left( \varphi_0 + \delta \right)}} {\widehat{\bf D}}_M^2 {\cal A}_1^{\ast}, \label{LinearizedFOMF}
\end{equation}
where 
\begin{equation}
{\widehat{\bf D}}_M = {\frac {\omega_0} {c}} \partial_t + c k_0 \partial_x, \qquad \quad {\widehat{\bf L}}_M = \Gamma {\Box} - \beta {\widehat{\bf D}}_M^2, \label{BasOperatMF}
\end{equation}
\begin{equation}
\beta = 4 \kappa \epsilon_0 {\left| {\cal E} \right|}^2, \qquad \qquad \Gamma = 1 - 4 \kappa \epsilon_0 c^2 B_0^2, \label{BetaCoeff}
\end{equation}
and $\delta$ is the phase of the constant complex amplitude ${\cal E}$. Applying again the averaging procedure described in the preceding Section, we neglect the fast oscillating term on the right-hand-side of Eq. (\ref{LinearizedFOMF}), and obtain the first-order solution 
\begin{equation}
{\cal A}_1 = {\cal B}_{+} {\rm e}^{i \varphi_1} + {\cal B}_{-} {\rm e}^{-i \varphi_1}. \label{FirstOrderSolMF}
\end{equation}
Here, ${\cal B}_{\pm}$ are supplementary constant complex amplitudes and $\varphi_1 = k x - \omega t$ is the wave phase. In addition, $\omega {\left( k \right)}$ is the wave frequency 
\begin{widetext}
\begin{equation}
\omega {\left( k \right)} = {\frac {1} {\Gamma + \beta \omega_0^2}} {\left[ \beta c^2 \omega_0 k_0 k + {\sqrt{\Gamma c^2 k^2 {\left( \Gamma + \beta \omega_0^2 \right)} - \Gamma \beta c^4 k_0^2 k^2}} \right]}, \label{OmegaFreqMF}
\end{equation}
\end{widetext}
which is a solution to the dispersion relation 
\begin{equation}
{\cal D}_M {\left( k, \omega \right)} = \Gamma {\Box}_k + \beta c^2 {\Box}_{0k}^2 = 0. \label{NonlinDisperMF}
\end{equation}

Since the main considerations and explicit calculations in higher orders are confined in the same mainstream of particulars as described in detail in Section \ref{sec:nlindisp}, we will skip them and will present only the final result. The sought for nonlinear amplitude equations for the slowly varying wave amplitudes ${\cal B}_{\pm}$ are 
\begin{widetext}
\begin{equation}
{\left[ {\widehat{\bf L}}_M + i {\left( \partial_{\omega} {\cal D}_M \right)} \partial_t - i {\left( \partial_k {\cal D}_M \right)} \partial_x \right]} {\cal B}_{+} = - 24 \kappa \epsilon_0 c^2 {\Box}_k^2 {\left| {\cal B}_{-} \right|}^2 {\cal B}_{+} + 2 \kappa \epsilon_0 c^2 {\frac {{\Box}_k} {{\Box}_{0k}}} {\left( 4 {\Box}_k^2 - 6 {\Box}_k {\Box}_{0k} + {\Box}_{0k}^2 \right)} {\left| {\cal B}_{+} \right|}^2 {\cal B}_{+}, \label{NonlinAmpEquMF}
\end{equation}
\begin{equation}
{\left[ {\widehat{\bf L}}_M - i {\left( \partial_{\omega} {\cal D}_M \right)} \partial_t + i {\left( \partial_k {\cal D}_M \right)} \partial_x \right]} {\cal B}_{-} = - 24 \kappa \epsilon_0 c^2 {\Box}_k^2 {\left| {\cal B}_{+} \right|}^2 {\cal B}_{-} - 2 \kappa \epsilon_0 c^2 {\frac {{\Box}_k} {{\Box}_{0k}}} {\left( 4 {\Box}_k^2 + 6 {\Box}_k {\Box}_{0k} + {\Box}_{0k}^2 \right)} {\left| {\cal B}_{-} \right|}^2 {\cal B}_{-}. \label{NonlinAmpEquMFMin}
\end{equation}
\end{widetext}
The above equations describe the nonlinear optical anisotropy of the polarized quantum vacuum and the presence of two intrinsic polarization states with different indices of refraction (nonlinear birefringence of the vacuum in an externally applied magnetic field). In other words, coherent interaction of photons in the pumping plane wave causes each of the induced waves to acquire an effective nonlinear refractive index. Obviously, the value of this index depends on the intensity of the pumping wave, as well as on the amplitude of the waves characterized by the two polarization states. After traversing a certain distance, a circularly polarized impacting wave is transformed into an elliptically polarized wave of the most general form. Polarization phenomena accompanying the interaction of two laser beams, which are due to scattering of light by light have been considered in the past\cite{Aleksandrov}.

\section{\label{sec:soliton}Nonlinear and Cnoidal Waves and Nonlinear Birefringence} 

It is instructive to transform Eqs. (\ref{NonlinAmpEqu}), (\ref{NonlinAmpEquMF}) and (\ref{NonlinAmpEquMFMin}) in a more convenient and familiar form. Let us deal with Eq. (\ref{NonlinAmpEqu}) first. Taking into account the translational invariance of our problem, we introduce new spatial variables according to the relation 
\begin{equation}
{\boldsymbol{\xi}} = {\bf x} - {\bf v} t, \qquad {\bf v} = {\frac {2 \alpha \omega_0^2} {1 + 2 \alpha \omega_0^2}} c {\widehat{\bf k}}_0, \label{TransVar}
\end{equation}
where ${\widehat{\bf k}}_0$ is the unit wave vector. This means that the new coordinate system is firmly fixed with an observer moving with a speed ${\bf v}$ (which for realistic impact wave intensities is a small number), as defined by the second of the above relations. In the new spatial variables defined above, the nonlinear amplitude equation (\ref{NonlinAmpEqu}) can be rewritten as 
\begin{widetext}
\begin{equation}
i {\left( \partial_{\omega} {\cal D} \right)} {\left[ \partial_t + {\left( {\bf v}_g - {\bf v} \right)} \cdot {\boldsymbol{\nabla}}_{\boldsymbol{\xi}} \right]} {\cal A} + {\left[ {\boldsymbol{\nabla}}_{\boldsymbol{\xi}}^2 - {\frac {2 \alpha \omega_0^2} {1 + 2 \alpha \omega_0^2}} {\left( {\widehat{\bf k}}_0 \cdot {\boldsymbol{\nabla}}_{\boldsymbol{\xi}} \right)}^2 - {\frac {1} {c^2}} {\left( 1 + 2 \alpha \omega_0^2 \right)} \partial_t^2 \right]} {\cal A} = - \sigma {\left| {\cal A} \right|}^2 {\cal A}. \label{NonlinAmplitudeEq}
\end{equation}
\end{widetext}
Here ${\bf v}_g = - {\left( {\boldsymbol{\nabla}}_k {\cal D} \right)} {\left( \partial_{\omega} {\cal D} \right)}^{-1}$ is the wave group velocity\cite{TzenovTUBES} and 
\begin{equation}
\sigma = 36 \kappa \epsilon_0 c^2 {\Box}_k^2, \label{NonlinCoeff}
\end{equation}
is the nonlinear coupling coefficient. 

Let us now assume that the vector potential (\ref{VecPotConfig}) depends on one spatial coordinate (say the $x$-coordinate) only. Thus, Eq. (\ref{NonlinAmplitudeEq}) acquires the form  
\begin{equation}
i {\cal D}_{\omega} {\left( \partial_t + \Delta v \partial_{\xi} \right)} {\cal A} + {\frac {1} {1 + 2 \alpha \omega_0^2}} {\widehat{\Box}}_p {\cal A} = - \sigma {\left| {\cal A} \right|}^2 {\cal A}, \label{NonlinAmplitMF}
\end{equation}
where 
\begin{equation}
{\widehat{\Box}}_p = {\boldsymbol{\nabla}}_{\boldsymbol{\xi}}^2 - {\frac {1} {c_1^2}} \partial_t^2, \qquad \quad c_1 = {\frac {c} {1 + 2 \alpha \omega_0^2}}. \label{OperVel}
\end{equation}
In addition, ${\cal D}_{\omega}$ is a short-hand notation of the derivative of the dispersion function (\ref{NonlinDisper}) with respect to $\omega$ and $\Delta v$ stands for the difference between the group velocity and the speed $v$ defined by Eq. (\ref{TransVar}) ${\left( \Delta v = v_g - v \right)}$. 

Here is the place to make a small diversion from the main exposition with an important comment. The nonlinear wave equation (\ref{NonlinAmplitMF}) governs the evolution of electromagnetic waves with a reduced speed of light $c_1$. For sufficiently small values of the parameter $\alpha$, the relative light velocity reduction $\Delta c / c = 2 \alpha \omega_0^2 {\left( 1 + 2 \alpha \omega_0^2 \right)}^{-1}$ is roughly proportional to the intensity of the pumping plane wave [cf. Eq. (\ref{AlphaCoeff})]. For example, at the maximum achievable value of the laser parameter $a_0 = 170$ at ELI-NP, this decrease is of the order of $\Delta c / c \sim 1.045 \cdot 10^{-10}$, while in the interstellar space at laser parameters of several orders of magnitude greater, the corresponding reduction in the speed of light is of the order of $\Delta c / c \sim 0.01 \; - \; 1.0 \%$. This implies that the effect mentioned above is expected to become more profound at a very high intensity of the impacting electromagnetic fields, and could be confirmed experimentally by monitoring and collecting of relevant data in a supernova explosion. A similar effect, although in a different context, was recently reported\cite{Franson,Jentschura}.

In order to find a stationary wave solution, we will take advantage of the universal ansatz 
\begin{equation}
{\cal A} {\left( \xi; t \right)} = B {\left( \zeta \right)} {\rm e}^{i \psi {\left( \xi; t \right)}}, \qquad \quad \zeta = \lambda \xi - u t, \label{Ansatz}
\end{equation}
where $\lambda$ is a constant, and $u$ is a constant phase velocity of the traveling wave, both to be specified in the sequel. Substituting the above expression for ${\cal A}$ into Eq. (\ref{NonlinAmplitMF}) and separating real and imaginary parts, we obtain two coupled differential equations relating the amplitude $B$ and the phase $\psi$. These can be written as 
\begin{widetext}
\begin{equation}
B {\widehat{\Box}}_p \psi + 2 {\left( \lambda \psi_{\xi} + {\frac {u {\dot{\psi}}} {c_1^2}} \right)} B^{\prime} - {\cal D}_{\omega} {\left( 1 + 2 \alpha \omega_0^2 \right)} {\left( u - \lambda \Delta v \right)} B^{\prime} = 0, \label{PhaseEq}
\end{equation}
\begin{equation}
{\left( \lambda^2 - {\frac {u^2} {c_1^2}} \right)} B^{\prime \prime} - {\left[ \psi_{\xi}^2 - {\frac {{\dot{\psi}}^2} {c_1^2}} + {\cal D}_{\omega} {\left( 1 + 2 \alpha \omega_0^2 \right)} {\left( {\dot{\psi}} + \Delta v \psi_{\xi} \right)} \right]} B = - \sigma {\left( 1 + 2 \alpha \omega_0^2 \right)} {\left| B \right|}^2 B, \label{AmplitEq}
\end{equation}
\end{widetext}
where the prime indicates differentiation with respect to the new variable $\zeta$, the over-dot implies differentiation with respect to time $t$ and the $\xi$-subscript denotes differentiation with respect to the $\xi$-variable. 

Let us now explore the travelling wave solutions possessed by the nonlinear wave equation (\ref{NonlinAmplitMF}). First, we consider the simple case, where $\psi {\left( \xi, t \right)} = \mu \zeta$. Moreover, $\mu$ is an arbitrary constant and without loss of generality, we can also assume that $\lambda = 1$. The phase equation (\ref{PhaseEq}) simply yields a relation between the known constants and the ones characterizing the travelling wave ansatz, which need to be determined additionally. Thus, we have
\begin{equation}
2 \mu - {\cal D}_{\omega} \gamma_u^2 {\left( 1 + 2 \alpha \omega_0^2 \right)} {\left( u - \Delta v \right)} = 0, \label{PhaseEq1}
\end{equation}
where 
\begin{equation}
\gamma_u = {\dfrac {1} {\sqrt{1 - {\dfrac {u^2} {c_1^2}}}}}, \label{Gammau}
\end{equation}
while the amplitude equation (\ref{AmplitEq}) can be rewritten as 
\begin{equation}
B^{\prime \prime} + \mu^2 B = - \sigma_1 B^3, \qquad \quad \sigma_1 = \sigma \gamma_u^2 {\left( 1 + 2 \alpha \omega_0^2 \right)}. \label{AmplitEq1}
\end{equation}
The above equation is the well-known Duffing equation, which in the case of positive $\sigma_1$ (which is the case, since $\sigma > 0$) has a simple solution expressed in terms of Jacobi cosine function. Assuming that $B^{\prime} {\left( \zeta = 0 \right)} = 0$, the exact solution of Eq. (\ref{AmplitEq1}) can be expressed as 
\begin{equation}
B = B_0 \cn {\left[ \nu {\left( \zeta + \psi_0 \right)}, k_e \right]}. \label{CnoidWave1}
\end{equation}
\begin{figure}
\begin{center} 
\includegraphics[width=8.0cm]{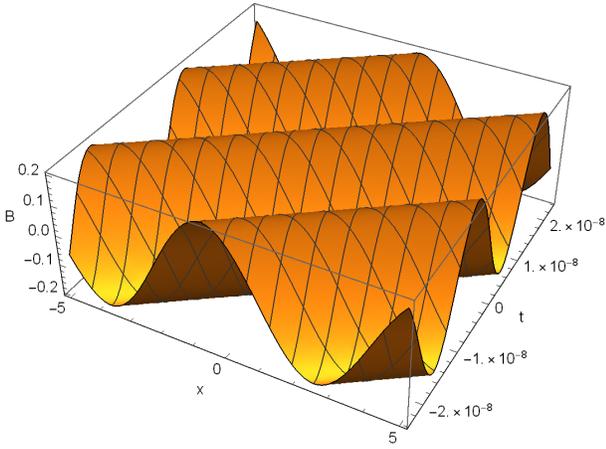}
\caption{\label{fig1:epsart} Evolution in space and time of the nonlinear wave amplitude ${\cal B}$ according to Eq. (\ref{CnoidWave1}). Shown here is the first harmonic with $\mu = 1$ for a typical laser intensity parameter $a_0 = 9.06$ of the pumping plane wave given by Eq. (\ref{PlaneWaveSol})}
\end{center}
\end{figure}
\begin{figure}
\begin{center} 
\includegraphics[width=8.0cm]{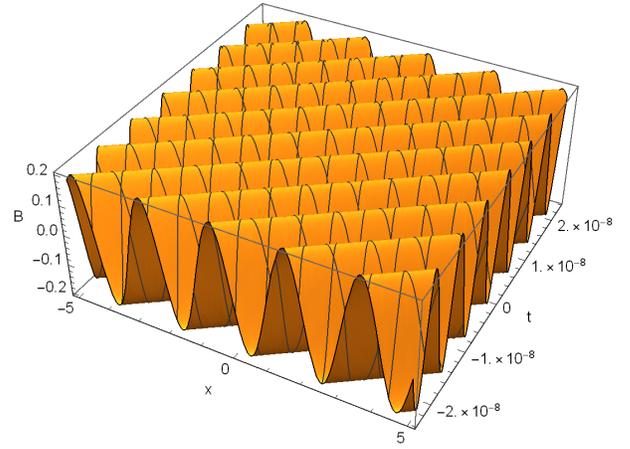}
\caption{\label{fig2:epsart} Evolution in space and time of the nonlinear wave amplitude ${\cal B}$ according to Eq. (\ref{CnoidWave1}). Shown here is the third harmonic with $\mu = 3$ for a typical laser intensity parameter $a_0 = 9.06$ of the pumping plane wave given by Eq. (\ref{PlaneWaveSol})}
\end{center}
\end{figure}
\noindent Here $\cn {\left( w \right)}$ denotes the elliptic Jacobi cosine function. In addition, $B_0 = B {\left( \zeta = 0 \right)}$, the frequency $\nu$ and the elliptic modulus $k_e$ are expressed as follows 
\begin{equation}
\nu = {\sqrt{\mu^2 + \sigma_1 B_0^2}}, \qquad k_e = {\dfrac {B_0} {{\sqrt{2 {\left( B_0^2 + {\dfrac {\mu^2} {\sigma_1}} \right)}}}}}. \label{Frequnuke1}
\end{equation}
and $\psi_0$ is an arbitrary phase. 

Next, we consider the general case of travelling wave solutions to the nonlinear wave equation, for which $\psi = \psi {\left( \zeta \right)}$. The phase equation (\ref{PhaseEq}) can be written as
\begin{equation}
\psi^{\prime \prime} B + 2 \psi^{\prime} B^{\prime} - \gamma B^{\prime} = 0, \label{PhaseEq2}
\end{equation}
where 
\begin{equation}
\gamma = {\cal D}_{\omega} \gamma_u^2 {\left( 1 + 2 \alpha \omega_0^2 \right)} {\left( u - \Delta v \right)}. \label{Constmu1}
\end{equation}
It possesses a first integral of the form 
\begin{equation}
C = B^2 {\left( \psi^{\prime} - {\frac {\gamma} {2}} \right)}, \label{FirstInteg}
\end{equation}
which after being substituted into the amplitude equation (\ref{AmplitEq}) yields the Ermakov-Pinney equation 
\begin{equation}
B^{\prime \prime} + {\frac {\gamma^2} {4}} B + \sigma_1 B^3 = {\frac {C^2} {B^3}}. \label{AmplitEq2}
\end{equation}
Multiplying both sides of Eq. (\ref{AmplitEq2}) by $B^{\prime}$ and integrating once, we obtain the first integral of the Ermakov-Pinney equation 
\begin{equation}
W = B^{\prime 2} + {\frac {C^2} {B^2}} + {\frac {\gamma^2} {4}} B^2 + {\frac {\sigma_1} {2}} B^4. \label{FIErmakPinney}
\end{equation}
Similar to the case worked out above, we assume that $B^{\prime} {\left( \zeta = 0 \right)} = 0$. Then, it can be verified by direct substitution that the Ermakov-Pinney equation (\ref{AmplitEq2}) possesses an exact solution in the form of a cnoidal wave 
\begin{equation}
B = {\sqrt{\frac {X_2 - k_e^2 X_3 \sn^2 {\left[ \nu {\left( \zeta + \psi_0 \right)}, k_e \right]}} {1 - k_e^2 \sn^2 {\left[ \nu {\left( \zeta + \psi_0 \right)}, k_e \right]}}}}. \label{CnoidWave2}
\end{equation}
The frequency $\nu$ and the elliptic modulus $k_e$ are given by the expressions 
\begin{equation}
\nu = {\sqrt{\frac {\sigma_1 {\left( B_0^2 - X_3 \right)}} {2}}}, \qquad k_e = {\sqrt{\frac {B_0^2 - X_2} {B_0^2 - X_3}}}. \label{Frequnuke2}
\end{equation}
The constants $X_2$ and $X_3$ entering the equations above can be expressed as follows 
\begin{equation}
X_{2,3} = - {\frac {1} {2}} {\left[ B_0^2 + {\frac {\gamma^2} {2 \sigma_1}} \mp {\sqrt{{\left( B_0^2 + {\frac {\gamma^2} {2 \sigma_1}} \right)}^2 + {\frac {8 C^2} {\sigma_1 B_0^2}}}} \right]}. \label{ConstX2X3}
\end{equation}
The analogy with the surface gravity waves on shallow water is quite interesting and impressive. Despite the fact that cnoidal waves have been proposed by Korteweg and de-Vries in 1895 and later modified by T. Benjamin, J. Bona and J. Mahony in 1972 in a different context\cite{Korteweg,Benjamin}, it is intriguing that they can play an important role in wave propagation in polarized quantum vacuum. 
\begin{figure}
\begin{center} 
\includegraphics[width=8.0cm]{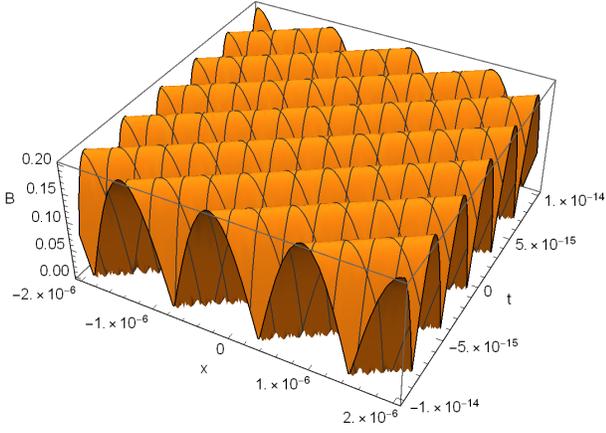}
\caption{\label{fig3:epsart} Evolution in space and time of the cnoidal wave amplitude ${\cal B}$ according to Eq. (\ref{CnoidWave2}). Shown here is the result for a typical laser intensity parameter $a_0 = 9.06$ of the pumping plane wave given by Eq. (\ref{PlaneWaveSol})}
\end{center}
\end{figure}

Let us now turn back to Eqs. (\ref{NonlinAmpEquMF}) and (\ref{NonlinAmpEquMFMin}) and perform the change of variables defined by Eq. (\ref{TransVar}). This time we define 
\begin{equation}
{\bf v} = {\frac {\beta \omega_0^2} {\Gamma + \beta \omega_0^2}} c {\widehat{\bf k}}_0. \label{TransVar1}
\end{equation}
In the new variables the coupled nonlinear equations (\ref{NonlinAmpEquMF}) and (\ref{NonlinAmpEquMFMin}) can be written as 
\begin{widetext}
\begin{eqnarray}
i {\cal D}_{M \omega} {\left( \partial_t + \Delta v_M \partial_{\xi} \right)} {\cal B}_{+} + {\frac {\Gamma^2} {\Gamma + \beta \omega_0^2}} {\widehat{\Box}}_M {\cal B}_{+} = - 24 \kappa \epsilon_0 c^2 {\Box}_k^2 {\left| {\cal B}_{-} \right|}^2 {\cal B}_{+} \nonumber
\\ 
+ 2 \kappa \epsilon_0 c^2 {\frac {{\Box}_k} {{\Box}_{0k}}} {\left( 4 {\Box}_k^2 - 6 {\Box}_k {\Box}_{0k} + {\Box}_{0k}^2 \right)} {\left| {\cal B}_{+} \right|}^2 {\cal B}_{+}, \label{NonlinAmpEquMF1}
\end{eqnarray}
\begin{eqnarray}
- i {\cal D}_{M \omega} {\left( \partial_t + \Delta v_M \partial_{\xi} \right)} {\cal B}_{-} + {\frac {\Gamma^2} {\Gamma + \beta \omega_0^2}} {\widehat{\Box}}_M {\cal B}_{-} = \nonumber 
\\ 
- 24 \kappa \epsilon_0 c^2 {\Box}_k^2 {\left| {\cal B}_{+} \right|}^2 {\cal B}_{-} - 2 \kappa \epsilon_0 c^2 {\frac {{\Box}_k} {{\Box}_{0k}}} {\left( 4 {\Box}_k^2 + 6 {\Box}_k {\Box}_{0k} + {\Box}_{0k}^2 \right)} {\left| {\cal B}_{-} \right|}^2 {\cal B}_{-}, \label{NonlinAmpEquMFMin1}
\end{eqnarray}
\end{widetext}
where now 
\begin{equation}
{\widehat{\Box}}_M = {\boldsymbol{\nabla}}_{\boldsymbol{\xi}}^2 - {\frac {1} {c_M^2}} \partial_t^2, \qquad \qquad c_M = {\frac {\Gamma c} {\Gamma + \beta \omega_0^2}}, \label{OperVel1}
\end{equation}
and the other notations are obvious and self-explanatory, in analogy to the case of free polarized vacuum, which has been just considered. Equations (\ref{NonlinAmpEquMF1}) and (\ref{NonlinAmpEquMFMin1}) possess a simple solution in the form of a plane wave ${\left[ {\cal B}_{\pm} = B_{\pm} {\rm e}^{i {\left( K \xi - \Omega_{\pm} t \right)}} \right]}$ with constant complex amplitudes $B_{\pm}$. The two frequencies $\Omega_{\pm}$ are solutions to the dispersion equations 
\begin{equation}
{\frac {\Gamma^2} {\Gamma + \beta \omega_0^2}} {\Box}_{\pm} \pm {\cal D}_{M \omega} {\left( \Omega_{\pm} - \Delta v_M K \right)} = \Sigma_{\pm}, \label{DisperBiref}
\end{equation}
where 
\begin{equation}
{\Box}_{\pm} = {\frac {\Omega_{\pm}^2} {c_M^2}} - K^2, \label{DisperBirefr}
\end{equation}
\begin{eqnarray}
\Sigma_{\pm} = - 24 \kappa \epsilon_0 c^2 {\Box}_k^2 {\left| B_{\mp} \right|}^2 \nonumber 
\\ 
\pm 2 \kappa \epsilon_0 c^2 {\frac {{\Box}_k} {{\Box}_{0k}}} {\left( 4 {\Box}_k^2 \mp 6 {\Box}_k {\Box}_{0k} + {\Box}_{0k}^2 \right)} {\left| B_{\pm} \right|}^2. \label{Sigmapm}
\end{eqnarray}
Since in the general nontrivial case both $\Sigma_{\pm}$ cannot be equal to zero, we would obtain $\Omega_{-} {\left( - K \right)} \neq - \Omega_{+} {\left( K \right)}$, that is $B_{-} \neq B_{+}^{\ast}$. This implies that the two wave modes ${\cal B}_{\pm}$ are independent and they propagate at different frequencies. This effect is usually called {\it nonlinear birefringence}. 

\section{\label{sec:conclude}Concluding Remarks} 

Starting from the basics, we have studied the implications of quantum corrections to classical electrodynamics and the propagation of electromagnetic waves and pulses. Used essentially is the property of the polarized vacuum, that the nonlinear wave equation for the electromagnetic vector potential possesses an exact solution in the form of a plane wave in both the case of free from external fields vacuum, as well as when a constant magnetic field is applied.

The initial nonlinear wave equation is solved perturbatively about the known exact plane wave solution. Following the elegant approach of the proto RG operator, we derive a nonlinear wave equation with a nonzero convective part (containing first-order spatial derivative with respect to the dependent variable) for the (relatively) slowly varying amplitude of the first-order perturbation. This equation governs the propagation of electromagnetic waves with a reduced speed of light. The reduction is roughly proportional to the intensity of the pumping plane wave. 

An external background magnetic field introduces spatial anisotropy coercing electromagnetic pulses to travel in the direction of the applied field. In analogy to the free vacuum case, the initial wave equation for the electromagnetic vector potential has been analysed in detail by following the same guidelines. In order to avoid insignificant complications in the specific calculations, we consider here the case of a circularly polarized initial pumping plane wave. The analysis of the general case of an elliptically polarized exact solution is also possible, but with it, the technical details conceal the physical nature of the problem. Repeating the proto RG procedure in a similar manner as has been done in the free vacuum case, we obtain a system of coupled nonlinear wave equations for the two slowly varying amplitudes of the first-order perturbation, which describe the two polarization states. 

The stationary (traveling) wave solutions of the proposed nonlinear wave equation with nonzero convective part for the case of a free polarized vacuum have been obtained. The slowly varying wave amplitude behaviour is shown to be similar to that of a cnoidal wave, known to describe surface gravity waves in shallow water. It has been demonstrated that the two wave modes describing the two polarization states are independent, and they propagate at different wave frequencies. This effect is usually called nonlinear birefringence.

\begin{acknowledgments}
We wish to express our gratitude to Prof. S.V. Bulanov and Dr. H. Kadlecova for many enlightening discussions and suggestions related to the topics touched upon in the article. 

The present work has been supported by Extreme Light Infrastructure -- Nuclear Physics (ELI-NP) Phase II, an innovative project co-financed by the Romanian Government and the European Union through the European Regional Development Fund. 
\end{acknowledgments}

\appendix

\section{\label{sec:mathieu}Perturbative Analysis of the Wave Analogue of Mathieu Equation (\ref{LinearizedFO})}

We consider the rapidly oscillating term on the right-hand-side of Eq. (\ref{LinearizedFO}) as perturbation and rewrite the latter accordingly 
\begin{equation}
{\widehat{\bf L}}_1 A_1 = - \epsilon \alpha {\left[ {\rm e}^{2i {\left( \varphi_0 + \delta \right)}} + {\rm e}^{-2i {\left( \varphi_0 + \delta \right)}} \right]} {\widehat{\bf D}}_1^2 A_1, \label{BasicAppend}
\end{equation}
where, once again, $\epsilon$ is a formal small parameter introduced to measure successive orders of magnitude in the perturbation expansion. Similar to Eq. (\ref{RGExpans}), we expand $A_1$ according to the expression 
\begin{equation}
A_1 {\left( x, z; t \right)} = \sum \limits_{n=0}^{\infty} \epsilon^n A_1^{(n)} {\left( x, z; t \right)}. \label{RGExpanApp}
\end{equation}
Obviously, the zero-order term $A_1^{(0)}$ is just the expression on the right-hand-side of Eq. (\ref{FirstOrderSol}). 

The first-order perturbation equation acquires the form 
\begin{eqnarray}
{\widehat{\bf L}}_1 A_1^{(1)} = - {\frac {{\Box}_k} {2}} {\left[ {\cal A} {\rm e}^{i {\left( \varphi_1 + 2 \varphi_0 + 2 \delta \right)}} \right.} \nonumber
\\ 
{\left. + {\cal A} {\rm e}^{i {\left( \varphi_1 - 2 \varphi_0 - 2 \delta \right)}} + c.c. \right]}. \label{FirOrdAppend}
\end{eqnarray}
Its solution can be written as 
\begin{equation}
A_1^{(1)} = - {\frac {{\Box}_k {\cal A}} {8 {\Box}_{0k}}} {\left[ {\rm e}^{i {\left( \varphi_1 + 2 \varphi_0 + 2 \delta \right)}} - {\rm e}^{i {\left( \varphi_1 - 2 \varphi_0 - 2 \delta \right)}} \right]} + c.c.. \label{FirOrdSolAppend}
\end{equation}
Finally, the second-order perturbation equation can be written as 
\begin{widetext}
\begin{equation}
{\widehat{\bf L}}_1 A_1^{(2)} = {\frac {{\Box}_k^2} {16 {\Box}_{0k}}} {\left[ {\rm e}^{2i {\left( \varphi_0 + \delta \right)}} + {\rm e}^{-2i {\left( \varphi_0 + \delta \right)}} \right]} {\left[ {\cal A} {\rm e}^{i {\left( \varphi_1 + 2 \varphi_0 + 2 \delta \right)}} - {\cal A} {\rm e}^{i {\left( \varphi_1 - 2 \varphi_0 - 2 \delta \right)}} + c.c. \right]}. \label{SecOrdAppend}
\end{equation}
\end{widetext}
It is straightforward to verify that resonant terms are missing on the right-hand-side of the above equation. Its solution contains only regular terms proportional to ${\rm e}^{i {\left( \varphi_1 \pm 4 \varphi_0 \pm 4 \delta \right)}}$ and their complex conjugated counterpart. This implies that the dispersion relation (\ref{NonlinDisper}) remains unaltered by the fast oscillating term in question, which completes the proof concerning the justification of the averaging procedure pursued in Section \ref{subsec:forder}. 
\bibliography{aipsamp}

\end{document}